\documentclass[acus]{jacow}
\usepackage{graphicx}
\usepackage{subfigure}
\usepackage{booktabs}
\usepackage{amsmath}
\usepackage{amssymb}
\usepackage{url}
\usepackage{threeparttable}
\usepackage{dcolumn}
\newcolumntype{d}{D{.}{.}{5.5}}
\newcolumntype{s}{D{.}{.}{1.2}}

\begin{document}

\title{EXPERIMENTAL RESULTS OF CRYSTAL-ASSISTED SLOW EXTRACTION AT THE SPS}


\author{M.A. Fraser\thanks{mfraser@cern.ch}, S. Gilardoni, B. Goddard, V. Kain, D. Mirarchi S. Montesano, S. Petrucci, S. Redaelli,\\R. Rossi, W. Scandale, L. Stoel\textsuperscript{1}, F.M. Velotti, CERN, Geneva, Switzerland\\
F. Murtas, INFN LNF, Frascati, Italy \\
F. Galluccio, INFN Sezione di Napoli, Naples, Italy \\
F. Addesa, G. Cavoto\textsuperscript{2}, F. Iacoangeli, INFN Sezione di Roma, Rome, Italy \\
\textsuperscript{1}also at Vienna University of Technology, Vienna, Austria\\
\textsuperscript{2}also at Sapienza University Roma, Rome, Italy\\
}

\maketitle

\begin{abstract}
The possibility of extracting highly energetic particles from the Super Proton Synchrotron (SPS) by means of silicon bent crystals has been explored since the 1990's. The channelling effect of a bent crystal can be used to strongly deflect primary protons and hence direct them onto an internal absorber or, with additional deflection elements, eject them from the synchrotron. Many studies and experiments have been carried out to investigate crystal channelling effects. As summarised in~\cite{herr, akbari, elsener1, elsener2}, extraction of 120 and 270~GeV proton beams has already been demonstrated in the SPS with dedicated experiments located in the ring. At present in the SPS, the UA9 experiment is performing studies to evaluate the possibility to use bent silicon crystals to steer particle beams in high energy accelerators~\cite{scandale1,scandale2,scandale3,scandale4,scandale5}. Recent studies on the feasibility of extraction from the SPS have been made using the UA9 infrastructure with a longer-term view of using crystals to help mitigate slow extraction induced activation of the SPS. In this paper, the possibility to eject particles from the SPS and into the extraction channel in Long Straight Section (LSS) 2 using the bent crystals already installed in the SPS is presented. Details of the concept, simulations and measurements carried out with beam are presented, before the outlook for the future is discussed.

\end{abstract}

\section{INTRODUCTION}

In view of the proposed Beam Dump Facility (BDF)~\cite{bdf} and continuing requests for higher beam intensities for Fixed Target physics, the application of bent crystals used in different configurations and modes of operation for slow extraction are being studied~\cite{fraser_ipac17, abtef}. Bent crystals offer promising solutions for reducing the activation of the SPS LSS2 extraction region that is induced by the small fraction of beam that unavoidably impinges the electrostatic septum (ZS) during the conventional resonant slow extraction process.  As a first step in 2016, the slow extraction of a 270~GeV proton beam into the TT20 extraction line towards the North Experimental Area of the SPS was demonstrated in dedicated Machine Development (MD) sessions using the extraction septa in LSS2 and a bent crystal, provided by the UA9 collaboration as part of their experimental installation in LSS5.

\section{EXTRACTION CONCEPT}

\begin{figure}[h]
   \centering
   \includegraphics[trim={310 25 5 20}, clip,width=60mm]{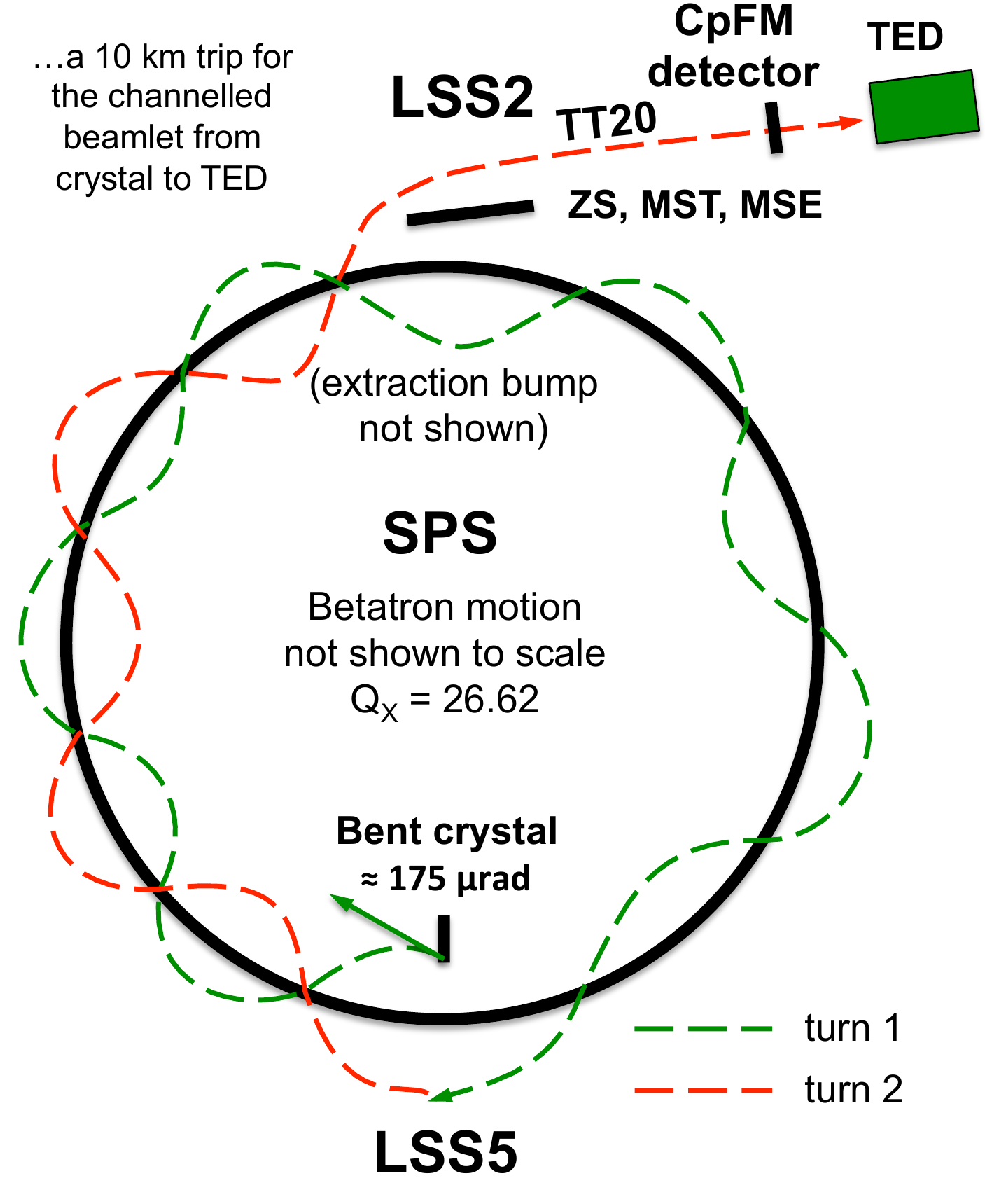}
  \caption{Schematic of the crystal-assisted extraction scheme.\label{schematic}}
  \end{figure}

The location of the bent crystals in LSS5 is 3.5~km from the slow extraction channel in LSS2, which makes the extraction process highly sensitive to the working point of the machine. The crystals are positioned on the inside of the ring and deflect inwards. Detailed studies~\cite{velotti_thesis} failed to identify a suitable working point that provides the required phase advance for the channelled beam to jump the wires of the electrostatic septum, on the outside of the ring, on the first turn. However, they did identify the potential of the operational Fixed Target working point ($Q_x = 26.62$, $Q_y = 26.58$) to extract the beam at the ZS on the second turn, with a fractional phase advance of $252^\circ$. The extraction scheme chosen is shown schematically in Fig.~\ref{schematic}, where the electrostatic and magnetic septa (ZS, MST and MSE) are shown, along with a dedicated Cherenkov for Proton Flux Measurement (CpFM) detector installed upstream of the extraction dump (TED) in the TT20 transfer line. The channelled beam performs almost 41 betatron oscillations before reaching the ZS.

With the machine configured to store a single bunch at constant energy, the transverse halo can be slowly and non-resonantly extracted as it diffuses into the crystal, is channelled and deflected into the extraction septa. The advantage of such an extraction concept for these first development tests is that the channelled beam passes the UA9 experimental area a second time, allowing the exploitation of the specialised beam diagnostics systems to verify the phase advance of the channelled beam in the absence of suitable systems in LSS2.

\begin{figure}[h]
   \centering
   \includegraphics[trim={20 20 10 285}, clip,width=83mm]{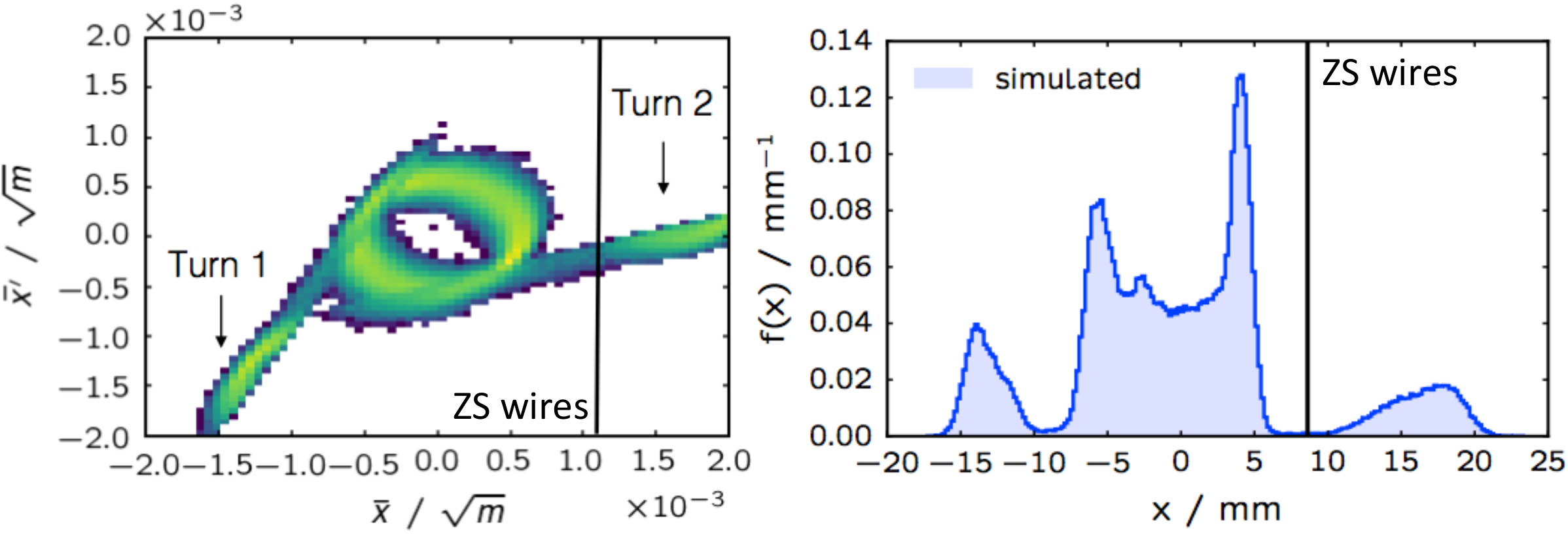}
  \caption{Beam distribution (hollow halo approximation) at ZS: normalised phase space (left) and horizontal projection (right)~\cite{velotti_thesis}. \label{x_px_at_ZS}}
  \end{figure}

In order to simulate the dynamics of the extraction, the SPS was implemented in MADX and particle tracking carried out in combination with the \texttt{pycollimate}~\cite{velotti_ipac} scattering routine, where the crystal interaction was modelled using single-pass UA9 measurement data~\cite{rossi}. More details of how the simulations were implemented are found in~\cite{velotti_thesis}. The presentation of the beam at the entrance to the ZS is shown in Fig.~\ref{x_px_at_ZS} after being tracked through the crystal and the SPS, where the distribution is approximated as a hollow halo beam to save computation time. In Fig.\ref{x_px_at_ZS} the crystal is aligned at $-6\sigma$ with a channelling angle of $-160~\mu$rad, where positive denotes a direction towards the outside of the ring. The simulations were used to design the LSS2 extraction bump, ensuring its closure, and to set the strengths of the extraction septa such that the channelled beam enters TT20 on the nominal trajectory. 

\vspace{-5mm}
\section{EXPERIMENTAL SETUP}

\begin{figure}[h]
   \centering
   \includegraphics[trim={35 30 10 325}, clip,width=82mm]{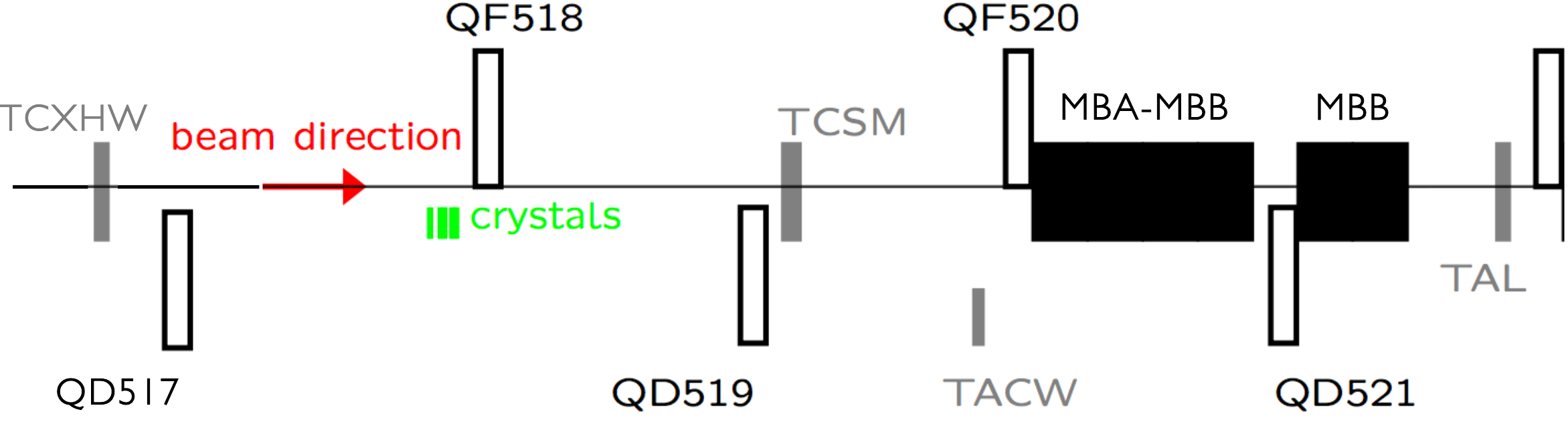}
  \caption{Schematic of the UA9 experiment's installation.\label{ua9_schematic}}
  \end{figure}

A schematic view of the UA9 installation is shown in Fig.~\ref{ua9_schematic} comprising two goniometers used to precisely align the crystals to the beam, several detectors used as beam diagnostics and to measure different observables, as well as absorbers and scrapers to intercept the channelled beam as well as to test the population of the halo at relevant phase advances with respect to the crystal.
The collimators and absorbers are named TCXHW, TCSM, TACW and TAL, and all of them are equipped with LHC Beam Loss Monitors (BLMs), which are significantly more sensitive than the standard SPS BLMs. The TCSM is an LHC prototype collimator with two horizontal jaws, composed of 1~m long blocks of graphite, and equipped with a Beam Position Monitor (BPM) at its entrance. The TACW is an old SPS collimator, which was equipped with a single 60 cm tungsten jaw to suit the needs of the UA9 experiment, used to stop the crystal-channelled beam during data taking. The experimental installation starts with the TCXHW, a 10 cm long double-sided tungsten scraper, and terminates with a station in the high dispersive area, which includes the TAL (10 cm long double-sided tungsten scraper) and a Roman Pot containing two Timepix high-precision pixel detectors~\cite{medipix_ref1,medipix_ref2}. In conjunction with UA9's absorbers and scrapers, the channelled beam could be stopped from circulating after a given number of turns and its position and transverse size measured on its return to LSS5, turn-by-turn. To directly detect the presence of channelled beam in the TT20 extraction line a dedicated CpFM detector~\cite{cpfm1,cpfm2} was installed that is capable of measuring single-particle events and therefore very low extraction rates.

\section{EXPERIMENTAL RESULTS}

\begin{figure}[h]
   \centering
   \subfigure[TACW IN.\label{tacw_in}]{
   \includegraphics[trim={15 15 20 95}, clip,width=39mm]{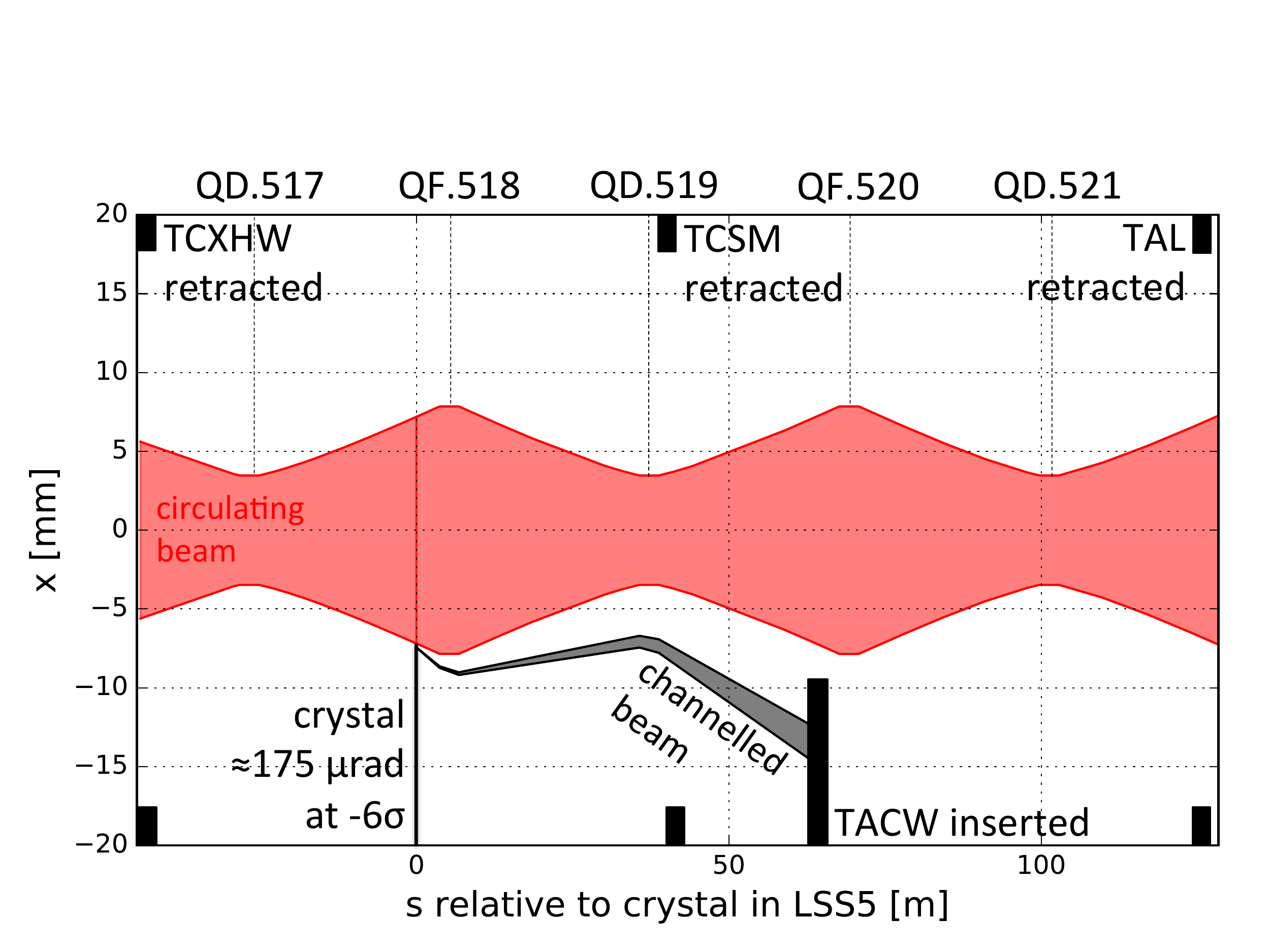}}
      \subfigure[TACW OUT.\label{tacw_out}]{
   \includegraphics[trim={15 15 20 95}, clip,width=39mm]{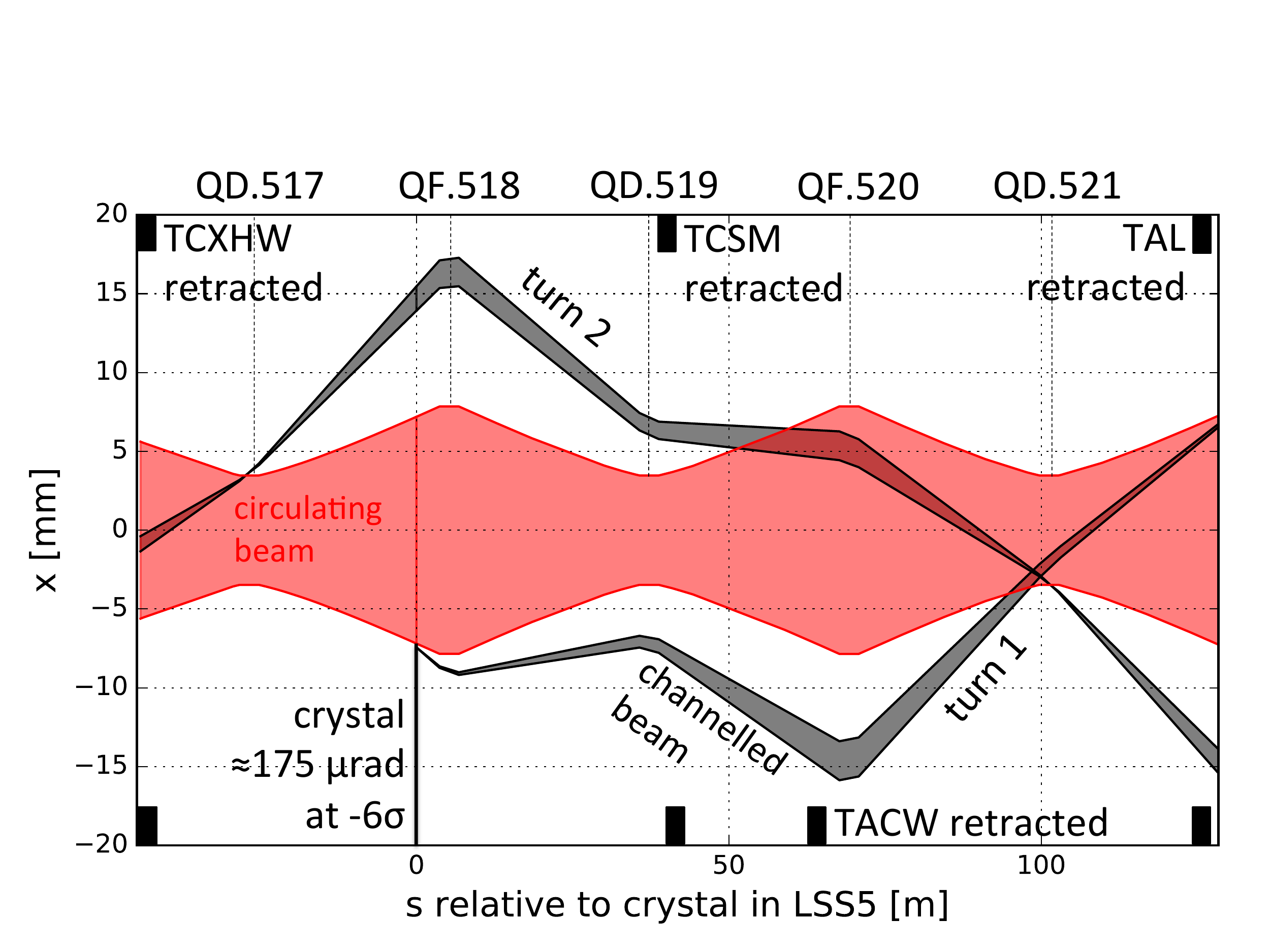}}
  \caption{LSS5 absorber configuration for extraction.~\label{tacw_position}}
  \end{figure}

A low intensity LHC-type single-bunch of $1.6\times10^{10}$ protons was used throughout the tests in order to guarantee the protection of the fragile wires of the ZS from damage. The effect of the imperfect closure of the LSS2 extraction bump was tested and shown not to significantly perturb the channelling efficiency. The BPM at the TCSM observed a 0.1~mm movement as the extraction bump was powered to its nominal value of 57 mm at the ZS, inducing a spike in the channelling rate and increasing BLM readings in LSS5. To avoid this type of dynamic effect it was decided to turn on the extraction bump before aligning the crystal with the beam. The alignment of the UA9 equipment was carried out as usual~\cite{ua9_procedure}.
\begin{figure}[h]
   \centering
   \includegraphics[trim={5 25 15 200}, clip,width=80mm]{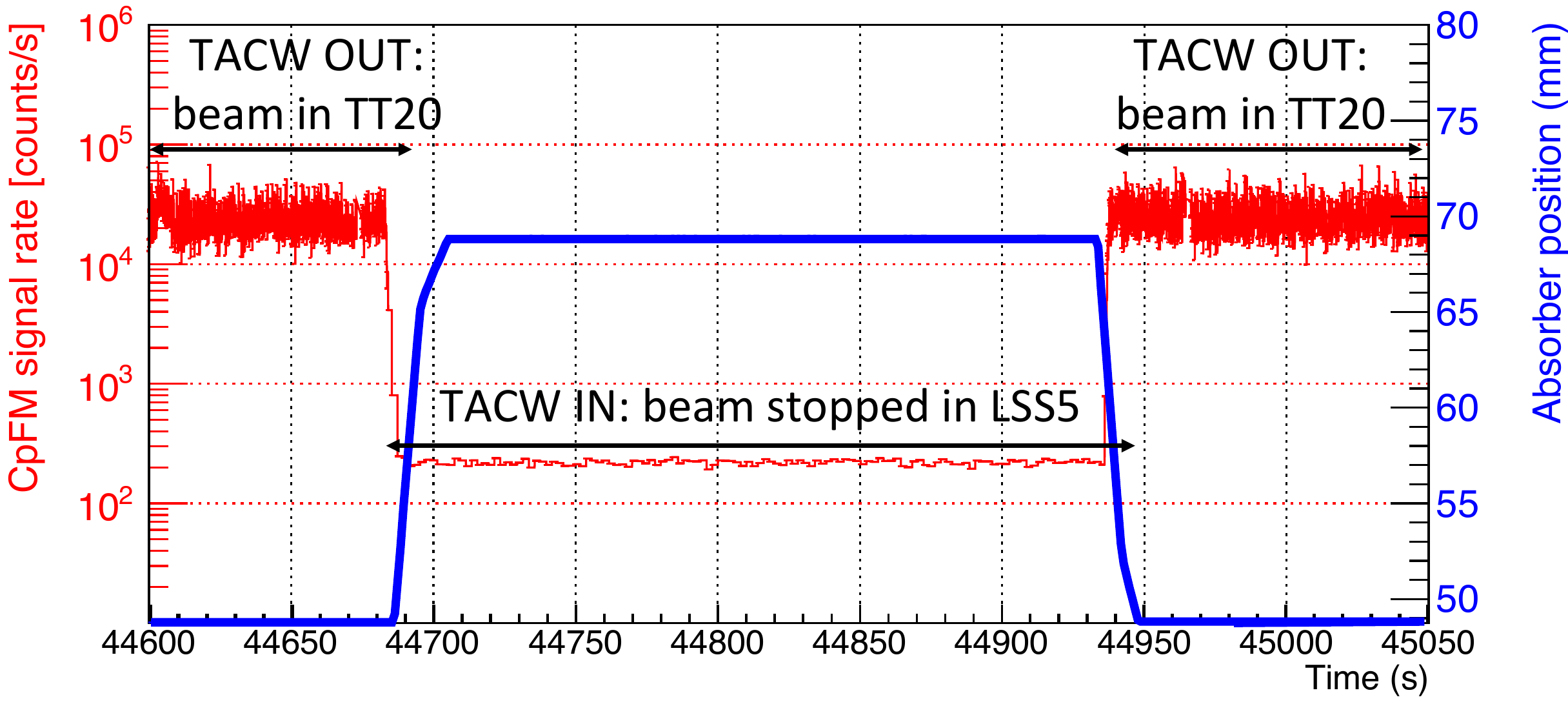}
  \caption{TT20 CpFM signal rate (red) vs. TACW position (blue)~\cite{cpfm2}. \label{CpFM_vs_TACW}}
  \end{figure}
The crystal was positioned at $6\sigma$ and aligned by scanning the angle of its goniometer and using scintillators and BLM loss levels as observables to determine the optimal orientation of the crystal. The bending angle of the crystal used was measured in a previous MD in 2016 as $175\pm2~\mu$rad, consistent with~\cite{scandale1}. Once the TACW absorber was retracted and the channelled beam was free to circulate, it was immediately detected by the CpFM in the TT20 extraction line. After small steering corrections the beam could be centred on the nominal beam axis at the CpFM. The extraction could be ceased by reinserting the TACW and stopping the channelled beam in LSS5 as shown in Figs.~\ref{CpFM_vs_TACW} and~\ref{tacw_position}.
  
\begin{figure}[h]
   \centering
   \includegraphics[trim={5 0 5 20}, clip,width=80mm]{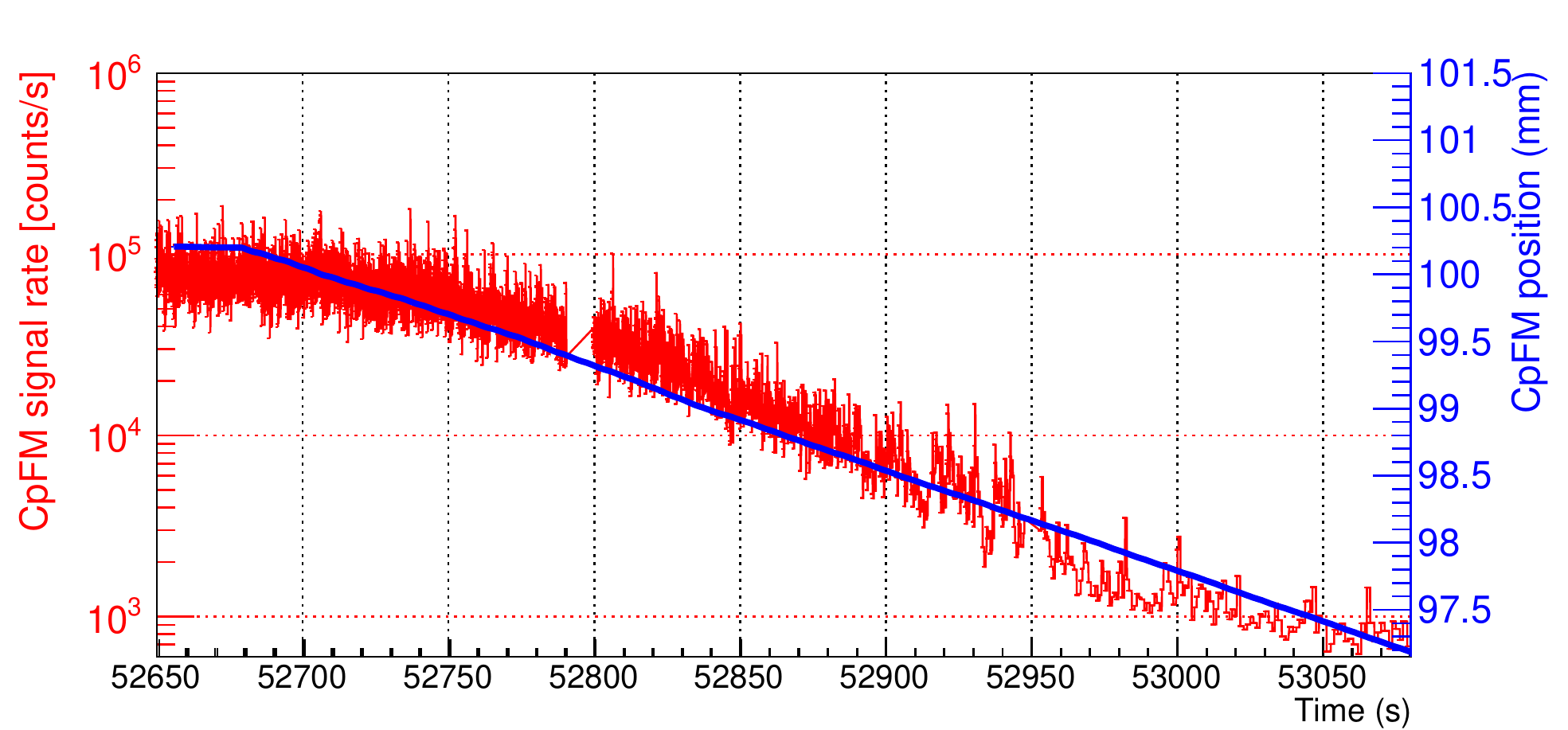} 
  \caption{TT20 CpFM signal (red) and position (blue) vs. time as it is moved out of the extracted beam~\cite{cpfm2}. \label{CpFM_scan}}
  \end{figure}

The horizontal beam size of the extracted beam at the CpFM was also measured at $\sigma = 0.6$~mm by scanning the quartz bar of the CpFM through the extracted beam in the horizontal plane, as shown in Fig.~\ref{CpFM_scan}. The signal reaches its maximum value when the quartz bar samples the entire beam. The calibration of the CpFM is on-going in order to quantify the exact extraction rate in terms of protons extracted per second.
\vspace{-0.5mm}

\begin{figure}[t!]
   \centering
   \subfigure[Extraction bump OFF.\label{medipix_1}]{
   \includegraphics[trim={10 5 5 260}, clip,width=82mm]{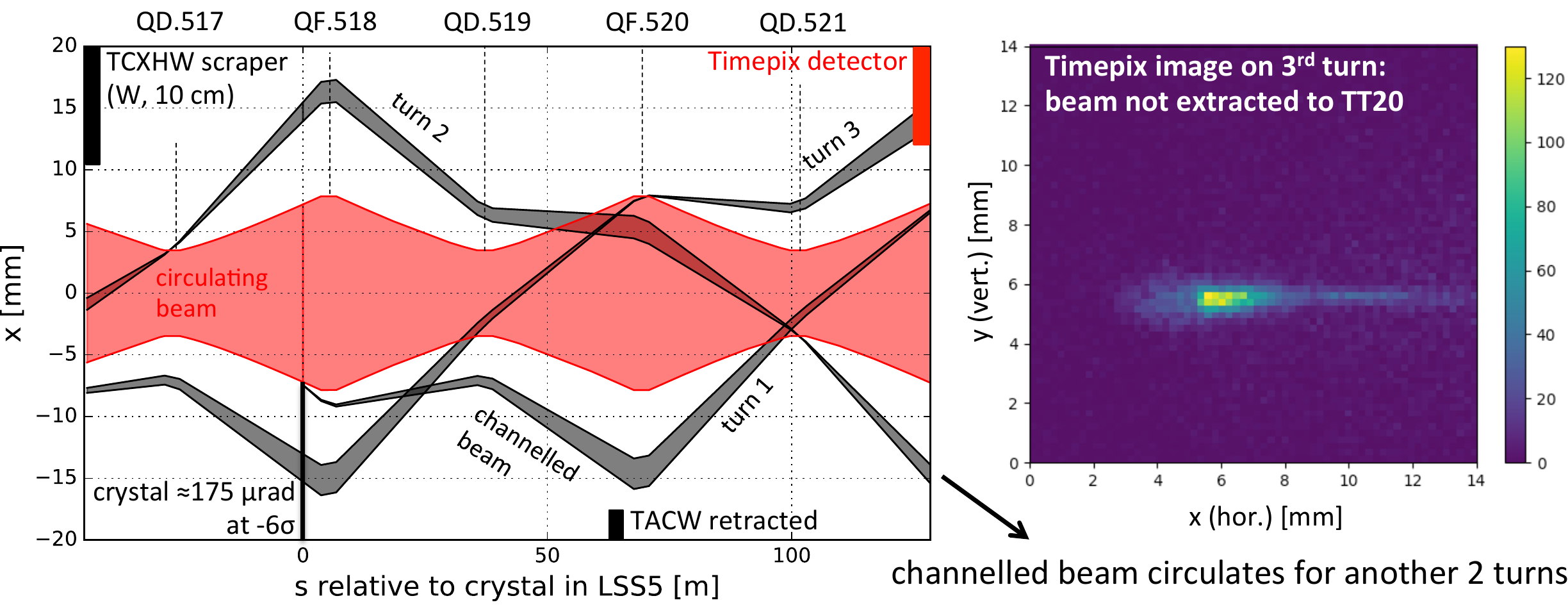}}
      \subfigure[Extraction bump ON.\label{medipix_2}]{
   \includegraphics[trim={10 10 5 260}, clip,width=82mm]{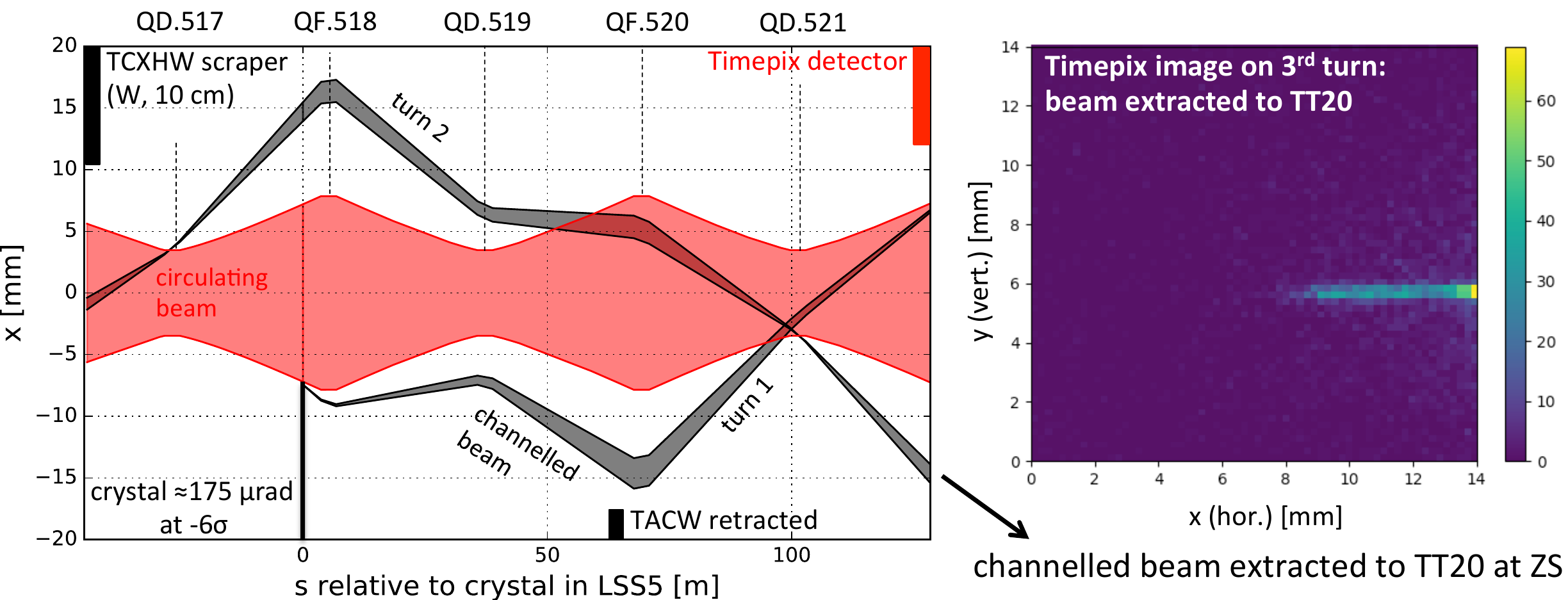}}
  \caption{Timepix images showing the disappearance of the channelled beam in LSS5 when extracted in LSS2.}
  \end{figure}

Further tests were made to guarantee that the extracted beam was indeed channelled by the crystal. These involved (i) inserting the outer TCSM jaw and TAL inner jaw to stop the channelled beam on the second turn in LSS5, also allowing beam profiles to be reconstructed, (ii) changing the angular alignment of crystal and (iii) adjusting the diffusion rate by exciting the beam with the transverse damper and checking the measured extraction rate on the CpFM. The aforementioned direct checks using the CpFM behaved as expected with the most elegant validation being an indirect measurement carried out with the Timepix detector. With the extraction bump turned off, and the TCXHW inserted to intercept the channelled beam from circulating after its fourth pass of LSS5, the channeled beam was imaged on the Timepix detector on its third pass of LSS5, as shown in Fig.~\ref{medipix_1}. When the extraction bump was turned on the channelled beam disappeared from the image as it was pushed over the septum wires of the ZS and extracted into TT20, as shown in Fig.~\ref{medipix_2}. The extraction was again confirmed by the CpFM.

\section{CONCLUSION AND OUTLOOK}

The non-resonant crystal-assisted slow extraction of a 270~GeV proton beam from the SPS towards the North Experimental Area has been demonstrated during dedicated MD tests in 2016. The low intensity beam was extracted from the halo of a circulating LHC-type single-bunch of $1.6\times10^{10}$ protons and its presence validated and characteristics probed with a dedicated CpFM detector in the extraction line. This is the first time in the SPS that a bent crystal has been used in conjunction with the extraction systems to bring the beam into a transfer line towards an experimental area. In light of future experimental requests for Fixed Target physics at 400~GeV, this is an important step in the application of bent crystals at the SPS for the mitigation of slow extraction induced activation.

The MD programme will continue in 2017 to increase the extraction rate, characterise the extracted beam, quantify and compare the losses for different extraction techniques, and calibrate the CpFM with a beam synchronous trigger to improve the signal-to-noise ratio. The application of bent crystals to shadow the wires of the ZS during a conventional resonant slow extraction is being actively studied~\cite{velotti_shadowing} and the installation of a dedicated crystal to test this proposal in MD sessions is presently being discussed.

\section{ACKNOWLEDGEMENTS}

We are grateful to the other members of the UA9 collaboration for their support and their contribution to the development of the experimental apparatus used in this study. This work has been partially supported by the ERC CoG CRYSBEAM G.A. 615089. We would also like to thank the SPS operation, beam instrumentation, septa and RF teams for their support, as well as the Experimental Areas and Physics co-ordinators for their flexibility and understanding in the scheduling of the dedicated MD sessions.

\raggedend

\end{document}